\documentclass[11pt,twocolumn]{article}
\usepackage[margin=1in]{geometry}

\title{A Survey of Distributed Intrusion Detection Approaches}
\author{Michael Treaster\\
National Center for Supercomputing Applications (NCSA)\\
University of Illinois\\
Email: treaster@ncsa.uiuc.edu}

\date{}

\begin{document}
\maketitle

\begin{abstract}

Distributed intrustion detection systems detect attacks on computer systems by analyzing data aggregated from distributed sources.  The distributed nature of the data sources allows patterns in the data to be seen that might not be detectable if each of the sources were examined individually.  This paper describes the various approaches that have been developed to share and analyze data in such systems, and discusses some issues that must be addressed before fully decentralized distributed intrusion detection systems can be made viable.    

\end{abstract}

\section*{Introduction}

Intrusion detection systems (IDS) have existed since the 1980's, ever since the rise of the Internet made it possible to attack computer systems from a remote terminal.  Although the first such systems operated independently on each machine on which they were installed, eventually the idea was proposed of aggregating IDS data from multiple machines in order to look for patterns across a network.  This can improve the system's ability to detect attacks that might otherwise be undetectable because each single host cannot does not have enough evidence to draw any conclusions.  

In general, distributed intrusion detection systems leverage some kind of single-node IDS software to monitor security events and collect data.  Therefore, research typically focuses more on the sharing, aggregation, and processing of this data from a variety of nodes rather than on the exact nature of the monitoring itself.  Existing approaches can be categorized along a variety of axes; here we examine data sharing, the nature of the data analysis, and security and trust features.

\section*{Data Sharing}

In a distributed IDS system, each agent shares its data with other agents in the system.  However, there are a wide variety of sharing schemes that have been developed.  These schemes can be viewed as a continuum, with centralized data reporting on one side and completely decentralized sharing on the other.  

The most extreme centralization is represented by systems in which a commercial vendor collects security information from a wide variety of customers, each running the vendor's agent software~\cite{deepsight, dshield}.  The vendor typically has multiple machines handling the data collection and analysis load that this widespread deployment incurs.  When the vendor detects a possible Internet-scale attack, customers receive alerts and advice from the professional security experts who manage the system.   This approach has two primary shortcomings.  First, the central management and processing of data represents a single point of failure or vulnerability.  Second, it results in a scalability bottleneck, and, due to the volume of incoming data, these systems often have slow response time to new threats.   

The most common distributed IDS approach is one in which all agents report data to a central server controlled at a domain or enterprise level~\cite{polla98hummingbird, snapp91dids, chatzigniannakis04distributed, huang99large, jackson91expert}.  This is fundamentally the same as in the previous centralization approach, but on a different scale, and this possesses most of the advantages and disadvantages of these larger-scale systems.  These are usually oriented towards enterprise security, and are generally unsuitable for use among independent peers on the Internet due to the central control.  

To address the scalability problem of a centralized system, many techniques use a hierarchical structure~\cite{balasubramaniyan98architecture, stanifordchen96grids, porras97emerald}.  Data is passed up a hierarchy tree and is processed at each level to search for intrusions and to reduce the amount of information that must be passed higher up the tree.  This helps address scalability and allows a system to be deployed across large enterprise-scale networks, but it limits the kinds of intrusions that can be detected at the highest levels.  This also helps address the single point of failure problem, since if a higher node in the hierarchy fails the lower tiers can typically continue to function, albeit with reduced detection capabilities.  

Between the hierarchical approach and the fully distributed approach lie projects such as \cite{gopalakrishna01framework}, which uses a hybrid hierarchical-distributed approach.  Each agent publishes ``interests'' to the network, which are distributed through a hierarchical structure.  Agents share data with other nodes who are interested, and all analysis occurs locally at the agent level.  

Instances of completely distributed solutions are much more rare and are much less well-developed.  Gossiping, multicast, or subscription-based data sharing techniques have been proposed~\cite{janakiraman03indra}, but none of these have yet been implemented in a distributed IDS system.  Other systems~\cite{vlachos04security} ignore the topic entirely or pass it off to the underlying peer-to-peer substrate.  Although these examples are still under development, they represent solutions that can be deployed on the Internet at large, independent of any central authority.

\section*{Nature of Data Analysis}

Although distributed IDS systems are usually independent of the techniques used to detect individual security events, the ways in which these security events are used can vary greatly.  Since most systems work in heterogeneous environments, and since the security relationship between, say, a port scan and a buffer overflow attack may not be obvious, how does a system turn event detection into a response?

Expert systems are a common approach~\cite{snapp91dids, jackson91expert}, relying on rule sets to process and respond to events.  These rules can attempt to define security policies, normal behavior, and/or anomalous behavior, and alerts or actions are generated based on how events match against the rules.  \cite{snapp91dids} attempts to map actions back to a particular human user, such that events can be correlated with the intentions of an individual.  

Many systems~\cite{barrus98distributed, stanifordchen96grids} use a threshold scheme.  Each security event increases a global alert level.  The amount of the increase can be based on any number of factors, such as the particular event that was observed and its relation to other events in time or space.  When the alert level exceeds a certain threshold, generic increased security measures are deployed, or an administrator is alerted.  Long periods of time without security events can cause the alert level to decrease.   

Augmented goal trees can be used to model intrusion possibilities~\cite{huang99large}.  As more states of the goal tree are fulfilled (based on data from the distributed agents), the system is able to anticipate and counter future stages of the intrusion.   An alternative graph-based approach in which connections between machines are logged and constructed into a graph of network activity has also been studied~\cite{stanifordchen96grids}.  These graphs are then analyzed by an expert system to detect possible intrusions.

\section*{Security and Trust}

Security and trust are crucial aspects of any distributed system.  However, in most proposed distributed IDS systems, however, these issues are given a much lower priority than other design considerations.  They address the possibility of a rogue agent or a denial of service attack on the system only in passing.  In all cases, a complete solution for trust and security is not provided, but sometimes a concrete solution to a limited aspect of the problem is presented.

One issue is that of message authentication, allowing agents to ensure that messages come from who they claim to come from.  Several systems~\cite{janakiraman03indra, vlachos04security} use signed messages, relying on a central certificate authority to generate the credentials.  This authority does not necessarily participate in the rest of the distributed IDS system.  Although this approach validates the source of a message and ensures that the contents have not been tampered with, it cannot protect a legitimate agent sending malicious data.

Smaller-scale, centrally controlled systems such as~\cite{polla98hummingbird} can rely upon a login mechanism, such as Kerberos.  Agents only acknowledge logged-in systems, providing a measure of trust to the validated agents.  This solution is only appropriate for systems with a central login authority, however.  This solution, like the signed message approach, is unable to protect against a legitimate agent sending malicious data.

The issue of trust can be left to individual agents in the system~\cite{porras97emerald}.  Each agent decides whether or not to trust higher level agents (``monitors'') in the system hierarchy.  The agent then subscribes to exchange information from those monitors it chooses to trust.  By aggregating and forwarding the data they receive from the lower level agents, the monitors are able to distribute data throughout the network.  It is not clear how a monitor protects against subscription by rogue agents which then feed it misinformation.

Denial of service attacks on agents can be detected using heartbeat signals~\cite{barrus98distributed}.   Each agent periodically sends a message to inform the rest of the system that it is functioning properly.  If other agents do not receive the heartbeat message on schedule, a denial of service attack is suspected and treated as another security event on the network.  

Beyond these initial approaches to security and trust, there has been little work in this area with regard to distributed intrusion detection systems, especially in systems with a centralized control component.  Most distributed IDS approaches ignore this topic entirely, but some list it among future work.   Several projects~\cite{janakiraman03indra, vlachos04security} suggest the possibility of using a ``web of trust'' among peers, but this approach has not yet been explored.

\section*{Future Directions}

We believe tolerance of misinformation is a key area in which to focus, due to the lack of attention that it has been given in previous work.  In existing systems, a rogue agent might easily corrupt the network by spreading incorrect data.  Systems must protect themselves against this type of attack.  Centrally managed systems can rely on having complete control over every agent in the network to protect themselves.  However, agents in a centrally managed system might be subverted, and fully decentralized systems cannot rely on this at all.  

One approach that has been suggested is to build a \emph{web of trust} between agents in the network.  As an agent reports information that is verified by others, the reputation of the agents is increased and it is trusted more in the future.  However, the system must protect against an agent adopting malicious behavior after building up a high level of trust.  This approach is closely related to several trust-oriented research endeavors~\cite{abdulrahman97distributed, chen03poblano, sniffen00trust}.   However, the details of a such a protocol have not yet been carefully specified.

\small
\bibliographystyle{abbrv}
\bibliography{treaster-dids-corr}

\end{document}